\documentclass[manuscript]{aastex}

\slugcomment{To appear in The Astrophysical Journal Letters}

\shorttitle{Variable circular polarization in LS Pegasi}
\shortauthors{Rodr\'\i guez-Gil et al.}

\begin{document}

\title{Evidence of magnetic accretion in an SW Sex star: discovery of variable
circular polarization in LS Pegasi\footnote{Based on observations made with the
William Herschel Telescope, operated on the island of La Palma by PPARC in the 
Spanish Observatorio del Roque de Los Muchachos of the Instituto de Astrof\'\i
sica de Canarias.}}


\author{Pablo Rodr\'\i guez-Gil and Jorge Casares}
\affil{Instituto de Astrof\'\i sica de Canarias, La
Laguna, E-38200, Santa Cruz de Tenerife, Spain}
\author{Ignacio G. Mart\'\i nez-Pais}
\affil{Instituto de Astrof\'\i sica de Canarias, La Laguna, E-38200, Santa Cruz de Tenerife, Spain; Departamento de Astrof\'\i sica, Universidad de La Laguna, La Laguna, E-38271, Santa Cruz de Tenerife, Spain} 

\author{Pasi Hakala}
\affil{Tuorla Observatory, University of Turku, FIN-21500, Finland}

\and

\author{Danny Steeghs}
\affil{Department of Physics and Astronomy, University of Southampton,
Southampton SO17 1BJ, UK}




\begin{abstract}
We report on the discovery of variable circular polarization in the SW Sex star
LS Pegasi. The observed modulation has an amplitude of $\approx 0.3~\%$ and a
period of 29.6 minutes, which we assume as the spin period of the magnetic white
dwarf. We also detected periodic flaring in the blue wing of H$_\beta$, with a
period of 33.5 minutes. The difference between both frequencies is just the
orbital frequency, so we relate the 33.5-min modulation to the \emph{beat}
between the orbital and spin period. We propose a new accretion scenario in SW
Sex stars, based on the shock of the disk-overflown gas stream against the white
dwarf's magnetosphere, which extends to the corotation radius. From this
geometry, we estimate a magnetic field strength of $B_\mathrm{1}\sim 5$--15 MG.
Our results indicate that magnetic accretion plays an important role in SW Sex
stars and we suggest that these systems are probably Intermediate Polars with
the highest mass accretion rates.
\end{abstract}



\section{Introduction}

SW Sex systems are a sub-class of nova-like cataclysmic variables (CVs) with peculiar
spectroscopic behavior \citep{thorstensen91}. They exhibit strong single-peaked
Balmer,
\ion{He}{1} and \ion{He}{2} emission lines which, with the exception of
\ion{He}{2}, remain largely unobscured during primary
eclipse (suggesting emission in vertical structures). The radial velocity curves of Balmer and \ion{He}{1} lines
show significant phase lags relative to photometric minimum, and display absorption components which show maximum depth at phase
opposite primary eclipse (\citealt{szkody90}).\par

Currently, none of the proposed models have succesfully explained all the SW Sex
peculiarities. The first model was proposed by
\citet*{honeycutt86}, who suggested the presence of an accretion disk wind.
\citet{williams89} invoked magnetic accretion,
with single-peaked lines arising in accretion columns above the orbital plane,
while \citet{hellier94} introduced a gas stream which overflows the disk. \citet{casares96} proposed an scenario in which matter is accreted on to the
magnetic poles of a synchronously rotating white dwarf. More recently,
\citet{horne99} introduced the idea of a disk-anchored magnetic propeller. The latest model was
proposed by \citet{groot99}, who invoked the presence of a strong shock that
occurs along the accretion stream trajectory, but inside the disk.\par

LS Peg (S193) is a non-eclipsing SW Sex system with an orbital period of
$P_{orb}\simeq$ 4.19 h (\citealt*{rodriguez98,martinez99,taylor99}). \citet{stockman92} did not detect a significant level of circular polarization
in the 5000--6000 \AA~band after averaging only two 8 minute exposures. The
reported polarization level was $-0.01 \pm 0.05$ \%.\par

\section{Observations and data reduction}

The spectropolarimetric observations of LS Peg were performed on the night of
1998 August 7 with the 4.2-m William Herschel Telescope (WHT) on La Palma and the ISIS triple spectrograph, using
the R600B grating and the $2048 \times 4096$ pixel$^2$ EEV10 CCD chip. A 1\arcsec~slit gave a spectral resolution of 1.44 \AA~ and the wavelength coverage was $\lambda\lambda$3720--5070. A quarterwave plate and a Savart calcite plate were
placed in the optical path, above and below the slit, respectively. The circular
polarization spectra were obtained as described by \citet{tinbergen92}. The Stokes parameter $V$ is contained in the intensity ratio of both spectra (ordinary and extraordinary), but it contains the response of the spectrograph. On the other hand, the ratio between the o- and e-spectrum in a given exposure is independent of the sky transparency. In order to eliminate any instrumental effects, we took a second exposure with the quarterwave plate rotated by 90 degrees. The spectrograph response remains unchanged whilst the contrast between the o- and e-ray is inverted relative to the first exposure. By comparing both exposures, the instrumental response is taken out. Each pair of exposures contains the level of polarization, $p_V=V/I$, where the Stokes parameter $I$ represents the total intensity.\par

A total of 38 spectra were obtained, covering approximately one orbital period
of the system. We used an exposure time of 250 seconds and spectra of a Cu-Ar
comparison arc lamp were taken regularly throughout the night in order to
achieve an optimal wavelength calibration. All the frames were de-biased,
flat-fielded and sky-subtracted in the standard way. Finally, the spectra were
optimally extracted \citep{horne86}. These reduction processes were performed
using IRAF\footnote{IRAF is distributed by the National Optical Astronomy
Observatories} routines. For wavelength calibration and most of subsequent
analyses we used the MOLLY package. A third-order polynomial was fitted to the
arc data, the $rms$ being always less than 0.1 \AA. The spectra were then
re-binned into an uniform velocity scale. We also observed a zero polarization
standard star to calibrate the results.

\section{Results}

\subsection{Periodic flaring in the wings of H$_\beta$}

Balmer lines in LS Peg show broad, asymmetric wings (see \citealt{martinez99};
\citealt{taylor99}), caused by a high-velocity emission component (an S-wave)
that crosses the line core from red to blue at $\varphi \sim 0$. At this point we have to note that we considered the red-to-blue crossing of the \ion{He}{2} radial velocity curve to be the instant of zero phase. This has been proved in eclipsing SW Sex stars (see e.g. \citealt{shafter88}; \citealt{still95}; \citealt*{smith00}). In figure~\ref{fig1} we present the trailed spectra of H$_\beta$, constructed from the normalized spectra. 
\placefigure{fig1}
If we closely inspect the wings of H$_\beta$, especially in the blue, we can easily detect the
presence of flares. This seems to indicate that the high-velocity components can
be modulated with a short period. We
measured the equivalent width (EW) of the blue wing of H$_\beta$ in all the
{\sl normalized} individual spectra, considering that the wing extends from --4000 km s$^{-1}$ to
--800 km s$^{-1}$ with respect to the line rest wavelength. With the aim of
finding periodicities, we computed a periodogram of the resulting EW curve,
which we show in the upper panel of figure~\ref{fig2}. 
\placefigure{fig2}
Two main peaks are clearly seen: one centred at the orbital frequency
($\nu_{orb}=5.72$ d$^{-1}$) and another one at $\nu_{EW}=43.0 \pm 3.0$
d$^{-1}$, which corresponds to a period of $P_{EW}=33.5 \pm 2.2$ minutes. We
must note that the error in $\nu_{EW}$ is half the FWHM of the corresponding
peak, as derived from a gaussian fit, so it is overestimated. In the lower panel
of figure~\ref{fig2} we show the EW data in phase with $P_{EW}$ after phase
folding the data into 20 phase bins. The modulation with this period is evident.

\subsection{Evidence of variable circular polarization}

We obtained the polarization spectra from the non-normalized spectra, as described in \S~1. As the total database
comprises 38 individual spectra, we have 19 spectra of circular polarization
with a time resolution of about 10 minutes. First, we derived the average
circular polarization level in the range $\lambda\lambda$3900--5070
(approximately the Johnson $B$-band) for each of the 19 polarization spectra,
including both the continuum and the lines. To find periodicities we constructed
the power spectrum of these data, which we present in figure~\ref{fig3}.
\placefigure{fig3}
There is a prominent peak centred at $\nu_p=48.7\pm3.1$ d$^{-1}$, which
corresponds to a period of $P_p=29.6\pm1.8$ minutes. After folding the
polarization data on this period, we find a clear modulation of
amplitude (peak-to-peak) $\sim0.3~\%$ (see figure~\ref{fig3}). To find whether the
observed polarization comes from the lines or from the continuum, we repeated
this analysis in selected regions of the continuum, finding exactly the same
periodicity and amplitude. These results indicate that the observed circular
polarization comes from the continuum.\par
The folded curve in figure~\ref{fig3} exhibits significant dispersion, that can be a consequence of short-scale source variability, so one can question the statistical significance of the result. We performed a sinusoidal fit to the folded curve, obtaining a reduced $\chi^2$ of 0.05~\%. The amplitude is 0.15~\%, that is, 3$\sigma$, so we can say that the result is statistically significant. Furthermore, achieving a polarization accuracy below 0.1~\% with  the ISIS spectrograph is possible. For instance, Dhillon \& Rutten (1995) measured a $0.11\pm0.02~\%$ linear polarization level in the SW Sex star V1315 Aql. They calibrated the results observing a zero-polarization star, finding no evidence for instrumental polarization at a level greater than 0.01~\%. 

\section{Discussion}

We have detected a clear EW modulation in the blue wing of H$_\beta$,
with a frequency of $\nu_{EW}=43.0 \pm 3.0$ d$^{-1}$. Similar flaring behavior has been
observed in H$_\alpha$, H$_\beta$, and \ion{He}{2}
$\lambda$4686 in the spectra of the SW Sex-type old nova BT Mon \citep{smith98},
with a period of $\simeq$~32 minutes, and more recently in the SW Sex star DW UMa \citep{smith00}, where a 30-minute period is found. This
flaring is typical of intermediate polar CVs (IPs, e.g. FO Aqr, \citealt{marsh96}; see also
\citealt{hellier99}).

We have also discovered variable circular polarization in LS Pegasi. It is the
first time that circular polarization is detected in an SW Sex system. Since the
measured polarization comes from the continuum, we tentatively attribute its
origin to cyclotron emission (see \S~4.2 for discussion). We associate the
period $P_p$ to the spin period of the white dwarf. This result seems to suggest
an ``IP-like'' structure in LS Peg. Of the known IPs, only three show circular
polarization: BG CMi \citep{penning86}, PQ Gem (\citealt{piirola93};
\citealt{rosen93}; \citealt{potter97}) and RX J1712.6-2414 \citep{buckley95}.
The reason why we only observe circular polarization in these IPs is that they
harbor the most magnetic white dwarves ($B_1\sim$ 2--8 MG, \citealt{wickra91}). The frequency of the EW modulation ($\nu_{EW}$) is slightly smaller than the
assumed spin frequency ($\nu_p$). The difference between them is $\nu_p -
\nu_{EW}=5.7$ d$^{-1}$ which coincides with the orbital frequency. Hence, we
identify $P_{EW}$ with the beat period (i.e. synodic) between the white dwarf
spin period and the orbital period.

\subsection{Accretion scenario}

\citet{groot99} reports the observation of a bright spot in SW Sex itself (when
the system was in low state) located well inside the outer disk radius, between
0.4--0.6 $R_{L_1}$, where $R_{L_1}$ is the distance from the inner Lagrangian
point ($L_1$) to the white dwarf. He suggested that this hot region is the
consequence of a strong shock between the gas stream and the disk, but he also
pointed out that the mechanism producing the shock is not clear. From the
results we present for LS Peg, we propose that this inner bright spot could be
produced by the shock of the disk-overflown gas stream against the white dwarf's
magnetosphere. The shock should occur close to the corotation radius (i.e. the
distance from the primary at which the Keplerian and white dwarf angular
velocities match, $r_{co}$) and above the disk plane. Current theories of
accretion disk-magnetic field interaction require the magnetospheric radius to
be nearly equal to the corotation radius (see e.g. \citealt{ghosh78}).\par
Let's assume that the shock takes place close to $r_{co}$. Then, $r_{shock}
\approx r_{co}=f R_{L_1}$, where $f \le 1$. The distance (expressed in units of
the orbital separation, $a$) between the white dwarf and the $L_1$ point is
given by $R_{L_1}/a=0.46(1+q)^{-1/3}$ \citep{warner76}, while the corotation
radius is:
\begin{equation}
r_{co}=\left[\frac{G M_1 P^2_1}{4 \pi^2}\right]^{\frac{1}{3}},   
\label{eq1}
\end{equation}
where $M_1$ is the mass of the primary and $P_1$ its spin period. From Kepler's
Third Law and assuming $r_{shock} \approx r_{co}$, we obtain:
\begin{equation}
P_1 \approx 0.31 f^{3/2} P_{orb}
\label{eq2}
\end{equation}
For LS Peg, $P_{orb}=0.174774$ d and $P_1=29.6$ min (from the circular
polarization modulation), so $f \approx 0.52$. This result is in good agreement
with the derived bright spot location in SW Sex, suggesting that the magnetic
shock mechanism could be also present in this system. Furthermore, we have
detected variable $B$-band circular polarization in SW Sex itself
(P. Rodr\'\i guez-Gil et al., in preparation) with a period of about 28 minutes. Entering this value
together with the orbital period in (\ref{eq2}), we get $f \approx 0.60$, which
is consistent with Groot's results.

\subsection{The magnetic field}

The magnetospheric radius in IPs, $r_0$, is usually given by $r_0 \approx 0.52
r_\mu$ \citep{ghosh79}, where $r_\mu$ is the Alfv\'en radius. Then,
\begin{equation}
r_0 \approx 0.52 r_\mu = 4.1 \mu^{4/7} {M_1}^{-1/7} {\dot{M}}^{-2/7},
\label{eq3}
\end{equation}
where $\mu$ is the magnetic moment of the primary, $M_1$ its mass, and $\dot{M}$
the mass transfer rate. As we mentioned above, the magnetospheric radius should
be comparable to the corotation radius, so let's assume that $r_0 \approx
r_{co}$. From (\ref{eq1}) and (\ref{eq3}), we obtain:
\begin{equation}
\mu \approx 6.49 \times 10^{-7} {M_1}^{5/6} {\dot{M}}^{1/2} {P_1}^{7/6}.
\label{eq4}
\end{equation}
Entering $P_1\simeq1800$ s, and $M_1=0.75~\mathrm{M}_\odot$ (a plausible mass
value for SW Sex systems), we finally get:
\begin{equation}
\mu \approx 1.8 \times 10^{25} {\dot{M}}^{1/2}.
\label{eq5}
\end{equation}
To derive the magnetic field of the white dwarf in LS Peg, we must know its
radius ($R_1$), since $B_1=\mu/{R_1}^3$. We can give an estimate of $R_1$ using
the mass-radius relation for white dwarfs with carbon cores of \citet{hamada61},
obtaining $R_1 \simeq 7.2 \times 10^8$ cm for a $M_1=0.75~\mathrm{M}_\odot$
white dwarf. On the other hand, we must enter a value for the mass accretion
rate, $\dot{M}$. From the results of \citet{echevarria94}, we can consider that
$\dot{M} \approx 10^{16}-10^{17}$ g s$^{-1}$ in SW Sex systems. Hence, the
magnetic field intensity of the white dwarf in LS Peg should be $$B_1 \approx
5-15~~\mathrm{MG}.$$
This range of values is consistent with what is found in the IPs which show
circular polarization. Nevertheless, we have to stress the fact that we see
modulated circular polarization in the $B$-band. In PQ Gem ($B_1 \approx 9$--21
MG, \citealt{potter97}), little circular polarization is detected in that band. As
cyclotron models predict, PQ Gem shows a level of circular polarization
increasing redwards from the $V$-band, so it is not completely clear that the
circular polarization we observe in LS Peg is related to continuum cyclotron
emission. On the other hand, we can not completely rule out this possibility due
to the fact that the polarization level in LS Peg is very low. It would be
crucial to perform longer wavelength circular polarimetry to determine whether
the circular polarization we see is of cyclotron emission origin.\par
We finally propose that magnetic accretion plays a fundamental role in the
behavior of SW Sex stars, and that they are IP systems with mass accretion rates
probably larger than in the rest of IPs in quiescence.

\acknowledgments
We thank Tom Marsh for allowing us to use his MOLLY package.

\clearpage



\figcaption[fig1.ps]{Trailed spectra of H$_\beta$ from the normalized individual
spectra (no phase binning has been applied). Contrast has been adjusted to
emphasize absorption ({\sl left}) or emission ({\sl right}). Black represents
emission and a whole orbital period is repeated for clarity.\label{fig1}}

\figcaption[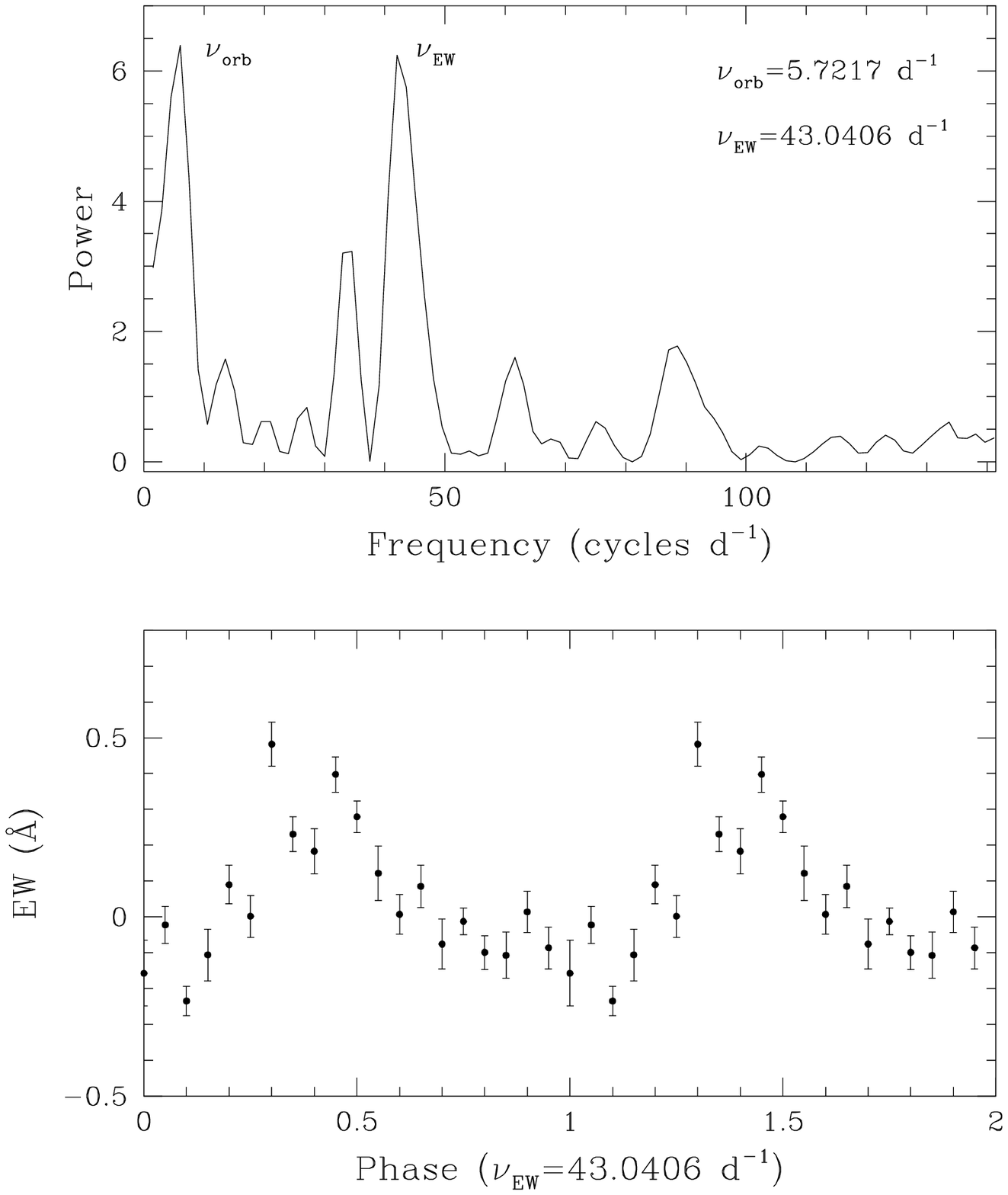]{{\sl Upper panel}: Power spectrum of the EW curve of
H$_\beta$ measured in the blue wing. There are two prominent peaks at the
orbital frequency and at $\nu\simeq43$ d $^{-1}$. {\sl Lower panel}: EW curve
folded on the period $P_{EW}=1/\nu_{EW}$ after averaging the data in 20 phase
bins. Zero phase is arbitrary and a complete period is repeated for
clarity.\label{fig2}}

\figcaption[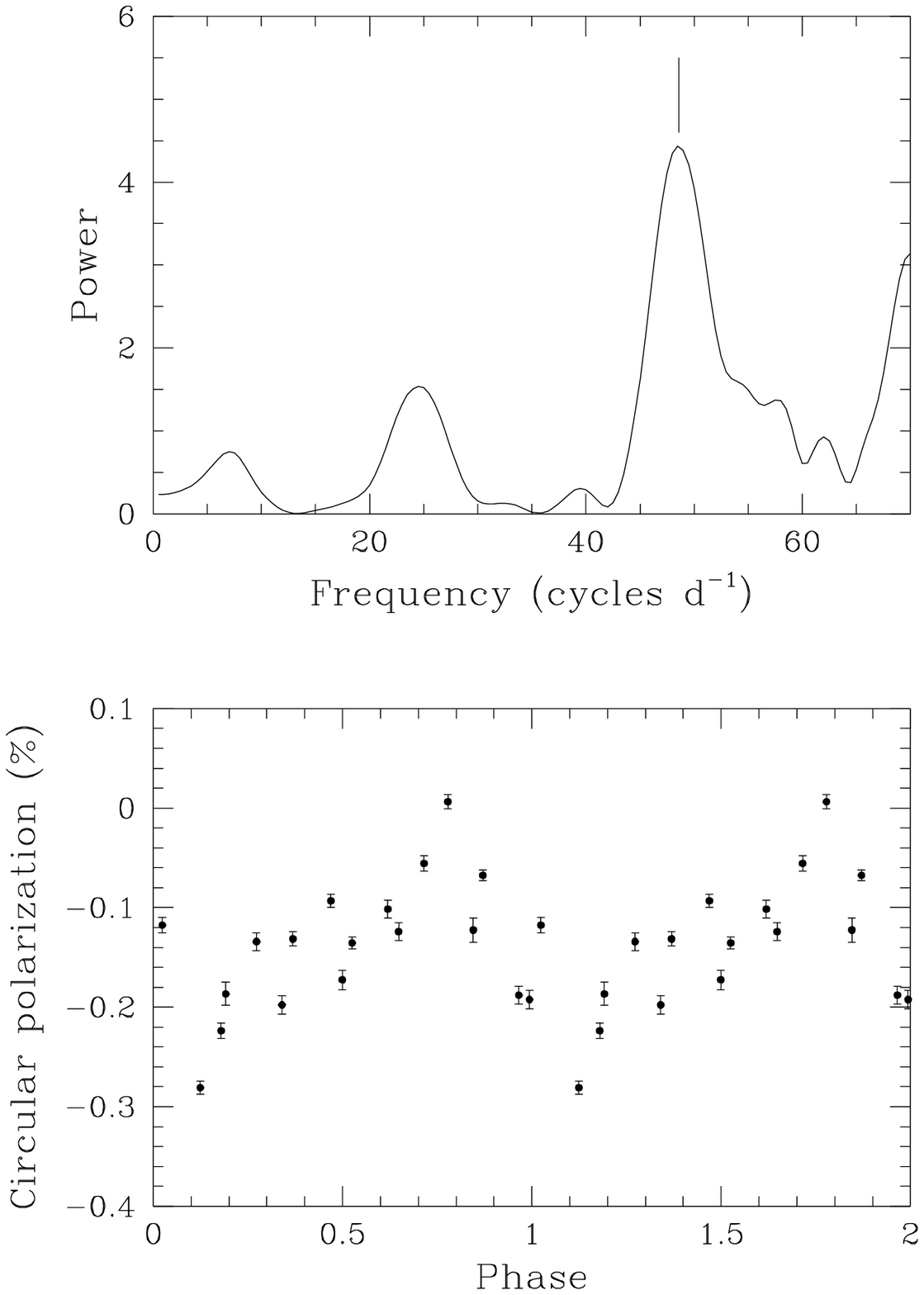]{{\sl Upper panel}: Power spectrum of the circular
polarization data in the $B$-band. The strongest peak at $\nu_p=48.7\pm3.1$
d$^{-1}$ is marked with a vertical line. {\sl Lower panel}: Circular
polarization level in phase with the period $P_p=1/\nu_p=29.6\pm1.8$ minutes.
Zero phase is arbitrary and a complete period is repeated for
clarity.\label{fig3}}





\clearpage

\end{document}